# Orthogonal spin arrangement as possible ground state of three – dimensional Shastry – Sutherland network in $Ba_3Cu_3In_4O_{12}$

O.S. Volkova[1], I.S. Maslova[1], R. Klingeler[2,3], M. Abdel-Hafiez[3], A.U.B. Wolter[3], B. Büchner[3], A.N. Vasiliev[1]

[1]Low Temperature Physics and Superconductivity Department, Moscow State University, Moscow 119991, Russia
[2]Kirchhoff Institute for Physics, University of Heidelberg, D-69120 Heidelberg, Germany
[3]Leibniz Institute for Solid State and Materials Research, IFW Dresden, D-01069 Dresden, Germany

The $Ba_3Cu_3In_4O_{12}$ stands for unique topology of the magnetic subsystem. It consists of rotated by 90° relative to each other "paper-chain" columns made of vertex-sharing $Cu^IO_4$ and $Cu^{II}O_4$ planar units. The overall pattern of the copper ions is that of a three-dimensional Shastry-Sutherland network. At high temperatures, the magnetic susceptibility follows the Curie-Weiss law with positive Weiss temperature indicating strong predominance of ferromagnetic coupling. At low temperatures, however, this compound reaches the antiferromagnetically ordered state and experiences non-trivial succession of two spin-flop and two spin-flip transitions reaching full saturation in modest magnetic fields. Here we show that the possible ground state in $Ba_3Cu_3In_4O_{12}$ can be three-dimensional orthogonal arrangement of the $Cu^{2+}$ ($S = 1/2$) magnetic moments forming three virtually independent antiferromagnetic subsystems. In this arrangement, favored by anisotropic exchange interactions, the quantum fluctuations provide the coupling between three mutually orthogonal magnetic subsystems resulting in an impressive "order by disorder" effect.

**INTRODUCTION**

Low-dimensional and frustrated magnetic systems have shown to attract attention by their rich variety of quantum ground states realized at low temperatures. Fascinating examples for exotic states are spin-liquids and various types of long-range collinear and non-collinear magnetic structures. Famous spin-liquids are the ancient Han-Purple pigment $BaCuSi_2O_6$[1], the spin-Peierls compound $CuGeO_3$[2], the Haldane chain $Y_2BaNiO_5$[3], and the Shastry-Sutherland network $SrCu_2(BO_3)_2$[4]. Representatives of non-collinear magnetic structures are helixes in quasi-one-dimensional quantum spin systems $LiCu_2O_2$[5], $LiCuVO_4$[6], and $Li_2ZrCuO_4$[7] where the arrangement of adjacent magnetic moments is a compromise between several isotropic superexchange interactions[8]. The pitch angle in these helixes depends on the ratio between magnetic nearest-neighbor and next-nearest-neighbor exchange interactions being close to 90° in a wide range of parameters[9]. There exists, however, another class of low-dimensional antiferromagnets where exact orthogonal arrangement of the magnetic moments is not governed by the isotropic superexchange, i.e. the tetragonal parent phases of electron-doped superconductors $R_2CuO_4$ (R = Nd,Pr,Sm). Here, the Cu magnetic moments in adjacent layers are oriented perpendicular to each other due to the anisotropic superexchange interaction[10]. For this class of materials, various sources of anisotropic contributions to the superexchange were investigated theoretically, including biquadratic superexchange[11] and pseudo-dipolar interactions[12].

The magnitude and the sign of the magnetic interactions in edge-sharing and corner-sharing geometries in cuprates are quite different[13]. The leading isotropic superexchange of a 180°-bond between two copper ions is strongly antiferromagnetic, while the leading order of a 90°-superexchange is ferromagnetic, and much weaker. In $Ba_3Cu_3In_4O_{12}$ the peculiar topology of the magnetic subsystem does not allow ascribing it to either of the above mentioned major classes. Similar to corner-sharing chains the $CuO_4$ units share only one vertex, but the Cu–O–Cu angle is close to 90° like in edge-sharing chains. According to our knowledge, neither any experimental study of the magnetic behavior nor calculations of magnetic exchange parameters in $Ba_3Cu_3In_4O_{12}$ have been undertaken so far.

$Ba_3Cu_3In_4O_{12}$ crystallizes in the tetragonal *I4/mcm* space group with lattice parameters $a$ = 12.121(3)Å, $c$ = 8.511(4)Å, $V$ = 1250(2)Å$^3$, $Z$ = 4, $c/a$ = 0.70[14]. The structure can be considered to be made of a perovskite-type network of $InO_6$ octahedra connected via vertices in three dimensions. Channels exist between these polyhedra which run parallel to the *c*-axis containing either columns of Ba atoms or chains of vertex-sharing $CuO_4$ square planes. Within a column there are two inequivalent crystallographic positions for copper ($Cu^I$ and $Cu^{II}$) in a 1:2 proportion. Each Cu–O chain is rotated by 90° in the *ab*-plane relative to its predecessor and the overall arrangement of the Cu–O chains is that of a square lattice, as shown in Fig. 1(left panel). The Cu–O sublattice of $Ba_3Cu_3In_4O_{12}$ consists of highly unusual chains of square planar units seen for the cuprates in this structure type only. The members of this family are $Ba_3Cu_3In_4O_{12}$[14], $Ba_3Cu_3Sc_4O_{12}$[15] and their solid solutions[15,16]. In terms of coordination polyhedra, buckled $Cu^IO_4$ square units form vertex-sharing chains in which each square shares four vertices. The chains are constructed such that each buckled $Cu^IO_4$ square is bridged by two $Cu^{II}O_4$ concave squares linked via opposite oxygen corners of the $Cu^IO_4$ square. Each successive pair of concave $Cu^{II}O$ squares is rotated by 90° with respect to its predecessor along the *c*-direction. This open linked chain, shown in Fig. 1(right panel), could be described as a paper-chain motif.

**EXPERIMENT**

Polycrystalline $Ba_3Cu_3In_4O_{12}$ was synthesized by means of a high temperature solid state reaction of stoichiometric amounts of $BaCO_3$, CuO, and $In_2O_3$. These chemicals were ground, pelleted and eventually fired in alumina crucibles at 850-950 °C in air for 72h with regrinding every 24h. Subsequently, the samples were quench-cooled in air to room temperature. The phase purity of the sample obtained was confirmed by powder X-ray diffraction data collected using a "Radian-2" diffractometer with Cu $K_\alpha$ radiation over a 2θ range of 20°-60°. The field and temperature dependences of the magnetization in $Ba_3Cu_3In_4O_{12}$ were measured in the temperature range 2–300 K and in magnetic fields up to 7T by means of a "SQUID-VSM" magnetometer (Quantum Design). Specific heat data were obtained by a relaxation technique in a PPMS (Quantum Design) in the temperature range 2–300 K on a sample with a mass of 12.7 mg.

**RESULTS**

The temperature dependence of the magnetic susceptibility $\chi$ = M/B in $Ba_3Cu_3In_4O_{12}$ taken at B = 0.1T is shown in Fig. 2(left panel). At low temperatures, the sharp peak implies the formation of long-range antiferromagnetic order at $T_N$ = 12.7K. In the magnetically ordered state the susceptibility drops by ~1/3 of its peak value which is typical for polycrystalline easy–axis antiferromagnets. At high temperatures, the susceptibility follows a Curie–Weiss behavior and the data can be approximated by the sum of a temperature independent term $\chi_0$ = –2.7×10$^{-4}$ emu/mol and the Curie-Weiss term $\chi_{CW}$ = C/(T–Θ). Fitting the data yields Θ = 52K and C = $N_A g^2 \mu_B^2 S(S+1)/3k_B$ = 1.36 emu·K/mol, where $N_A$, $\mu_B$ and $k_B$ are the Avogadro, Bohr and Boltzmann numbers. The value of the Curie constant gives a g-factor g = 2.2. The obtained value of $\chi_0$ is somewhat lower than the summation of Pascal's constants for the ions constituting $Ba_3Cu_3In_4O_{12}$: –3.3×10$^{-4}$ emu/mol[17], which is probably due to an additional temperature independent van Vleck contribution of $Cu^{2+}$ ions. The positive value of the Weiss temperature Θ indicates the predominance of ferromagnetic coupling at high temperatures. However, there are significant deviations between the Curie-Weiss fit and the experimental data well above the actual long-range ordering temperature $T_N$ as well as of the Weiss temperature Θ (see Inset of Fig.

2(left panel)). These deviations indicate the increasing relevance of antiferromagnetic correlations upon cooling which eventually yield the evolution of long-range antiferromagnetic order at $T_N$.

The field dependences of the magnetization M taken at selected temperatures in both the paramagnetic and the magnetically ordered state of $Ba_3Cu_3In_4O_{12}$ are shown in Fig. 2(right panel). At T = 100K, the magnetization depends practically linear on magnetic field up to 7T. Upon lowering the temperature, however, these curves significantly deviate from linearity showing pronounced right-bending thereby qualitatively indicating a ferromagnetic-type behavior. In order to get further insight into the evolution of the spin correlations above $T_N$ we have analyzed the M(B) curves using a modified Brillouin function $M(B) = M_S \cdot tanh(NM_SB/k_BN_AT)$[18]. Here, $M_S$ is the saturation moment of the nonlinear contribution to M(B) which is described by the $B_{1/2}$ Brillouin function, N is the number of ferromagnetically correlated spins at a given temperature T. Fitting the data with $M_S$ independent on the temperature and N(T) being the variable parameter provides a good description of the experimental data. At lowering temperature, the number N of correlated spins increases twice from 100K to 20K, but not diverge approaching $T_N$. The increasing number of correlated spins upon cooling implies larger ferromagnetic correlations at low temperature. Considering on the other hand the vicinity of long range antiferromagnetic ordering at $T_N$, the data indicate either one- or two-dimensional ferromagnetic fluctuations emerging upon cooling while additional antiferromagnetic coupling yield long range magnetic order below $T_N$. At the transition into the magnetically ordered state the M(B) curves acquire qualitatively new features at moderate magnetic field. At lowest temperature, T = 2 K, the magnetization M reaches saturation of about 3.0 $\mu_B$/f.u. (cf. with expected $M_S$ = 3.3 $\mu_B$/f.u. for g-factor g = 2.2) in a modest magnetic field of about 5.2T.

The transition to the magnetically ordered state of $Ba_3Cu_3In_4O_{12}$ is quite pronounced in specific heat measurements, as shown in Fig. 3(left panel). The main feature observed in Cp(T) at B = 0 is the λ-type anomaly at $T_N$ indicating a second order phase transition. At $T_N$, the experimentally observed specific heat changes by $\Delta Cp$ = 5.15 J/mol·K. Even when possible effects of fluctuations on the anomaly size are neglected, this value is by far, i.e. ~7.3 times, smaller of the expected jump size $\Delta Cp_{magn} = 5nRS(S+1)/[(S+1)^2+S^2]$ = 37.4 J/mol·K in the mean field model, with the number of magnetically active ions per formula unit n = 3, and the gas constant R = 8.314 J/mol·K. In magnetic fields the λ-type anomaly progressively smears out and shifts to lower temperatures, so that no singularity at the phase transition can be identified for T ≥ 2K and B ≥ 5T.

The inset to Fig. 3(left panel) presents specific heat data measured up to room temperature. For $Ba_3Cu_3In_4O_{12}$ the Dulong-Petit value reaches 3Rz = 549 J/mol·K, with the number of atoms per formula unit z = 22. In the Debye model about 0.95 of this value is reached at the Debye temperature $\Theta_D$. In $Ba_3Cu_3In_4O_{12}$ this temperature amounts to ~280K. This allows approximating the phonon contribution to the heat capacity at low temperatures $C_{ph} = \beta T^3$ function with $\beta = 1.9 \cdot 10^{-3}$ J/mol·K$^4$.

The total entropy released at T ≤ $T_N$ is comprised of various contributions and amounts to $\Delta S_{total}$ = 6.65 J/mol·K. The phonon part in this temperature range is $\Delta S_{phonon}$ = 1.35 J/mol·K in accordance with the estimation of the Debye temperature. To separate the magnon contribution one has to calculate the Fisher's specific heat $d(\chi_{//}T)/dT$, where $\chi_{//}$ is the longitudinal magnetic susceptibility[19]. In polycrystalline easy-axis antiferromagnets the magnetic susceptibility $\chi = \chi_{//}/3 + 2\chi_{\perp}/3$, where the transverse magnetic susceptibility $\chi_{\perp}$ is temperature independent at T ≤ $T_N$. It allows extracting the longitudinal susceptibility $\chi_{//}$ from experimental data $\chi(T)$ in $Ba_3Cu_3In_4O_{12}$ and, therefore, calculate the magnon contribution in the magnetically ordered state. The Fisher's specific heat $d(\chi_{//}T)/dT$ normalized to the experimental data of Cp at $T_N$ = 12.7K with account of phonon's contribution is shown in Fig. 3(right panel).

Evidently, the magnon contribution, as expected for a 3D antiferromagnet, is $\sim \alpha T^3$ with $\alpha = 3.98 \cdot 10^{-3}$ J/mol·K$^4$. Thus, we arrive at $\Delta S_{magnon} = 2.67$ J/mol·K. This value is ~0.154 of the one estimated within the mean field model $nR\ln(2S+1) = 17.29$ J/mol·K and is in good agreement with the value of the jump size (~0.137) in the specific heat at $T_N$. The extraction of both phonon $\Delta S_{phonon}$ and magnon $\Delta S_{magnon}$ contributions from the total entropy $\Delta S_{total}$ indicates that there is an extra contribution to the entropy $\Delta S_{Sch} = 2.63$ J/mol·K. The specific heat corresponding to this extra contribution $\Delta C_{Sch}$ is shown in the Inset of Fig. 3(right panel). Basically, this non-monotonous contribution is of a Schottky-type, but a fit to the experimental data using $C_{Sch} \sim R(\Delta/T)^2 \exp(\Delta/T)/(1+\exp(\Delta/T))^2$ with $\Delta \sim 10$K is rather poor due to numerous approximations needed to extract this term. Nevertheless, it can't be attributed to impurities since no indications for that are seen in magnetic susceptibility data.

**DISCUSSION**

At first glance, Ba$_3$Cu$_3$In$_4$O$_{12}$ hardly seems anything but a three-dimensional easy-axis antiferromagnet. The predominance of ferromagnetic coupling at elevated temperatures in the paramagnetic state is in accordance with the assumption of the leading role of a ~ 90° bond Cu$^I$-O-Cu$^{II}$ nearest-neighbor superexchange interaction $J < 0$ (cf. Fig. 1). However, strong deviation from the "ferromagnetic" Curie-Weiss law seen in this compound at low temperatures can be expected taking into account well-established frustrating influence of the next-nearest-neighbor superexchange interaction $J_1'' > 0$ through Cu$^{II}$–O–O–Cu$^{II}$ bonds[20]. Depending on the frustration ratio $\alpha = J_1''/J$, the combined action of (J, $J_1''$) interactions may result in the formation of either helix structures with various pitch angles at $\alpha \geq 0.25$ or ferromagnetic chains at $\alpha \leq 0.25$[21]. Moreover, the ferromagnetic chain arrangement may also survive at $\alpha \geq 0.25$ if it is stabilized by interchain interactions $J_2''$[22].

The naive scenario for the long-range magnetic order in Ba$_3$Cu$_3$In$_4$O$_{12}$ is, therefore, that of ferromagnetic "paper-chain" columns which are coupled antiferromagnetically. In this simplified model $J_1' = J_1'' = J_1$ and $J_2' = J_2'' = J_2$. Then, in accordance with the mean field formula, $g\mu_B B_S = 2zJ_2S$, for a spin $S = 1/2$, a g–factor $g = 2.2$ and the number of interchain nearest-neighbors $z = 4$ the value of the determined saturation field $B_S = 5.2$T (~3.5K) defines the interchain exchange interaction parameter $J_2 \sim 2$K. Through the simple expression $T_N = (|J| \times J_2)^{1/2}$ one can estimate the nearest-neighbor intrachain exchange interaction parameter $J = –84$K. In the mean – field theory the summation of exchange interaction parameters $J_i$ is presumed for the Weiss temperature $\Theta = \Sigma_i z_i S(S+1)|J_i|/3$. Taking the intrachain nearest-neighbors number $z = 2$ and the intrachain next-nearest-neighbors number $z=4$ one can get the value of the next-nearest-neighbor exchange interaction parameter within the chains $J_1 = 8$K. This, in turn, gives the frustration ratio $\alpha = J_1/J \sim 0.1$ in accordance with the postulated scenario. Of course, significantly elaborated formula can be used to estimate the exchange interaction parameters J, $J_1$ and $J_2$ in quasi-one-dimensional systems within a modified mean field theory[23,24]. For given values of the saturation field $B_S \sim 5.2$T and the Neel temperature $T_N = 12.7$K all of them result in even smaller values of $J_1/J$ and $J_2/J$ ratios.

However, the apparent triumph of the mean field theory for the case of antiferromagnetically coupled ferromagnetic chains is quite doubtful, if to compare Ba$_3$Cu$_3$In$_4$O$_{12}$ with any other quasi-one-dimensional compound and consider the minute details of the available data. Taking as representative the helix chain compound LiCuVO$_4$[25] and the ferromagnetic chain compound Li$_2$CuO$_2$[22] one may notice the striking disparity with Ba$_3$Cu$_3$In$_4$O$_{12}$ regarding interrelations of the magnetic ordering temperature $T_N$, the saturation field $B_S$ and the exchange interaction parameters $J_i$, as given in Table 1. Evidently, the magnetic field $B_S$ necessary to saturate the magnetization in both helix and ferromagnetic chain compounds is by an order of magnitude larger than in paper-chain compound despite the relatively high magnetic ordering temperature $T_N$ in the latter.

The low value of the saturation field in $Ba_3Cu_3In_4O_{12}$ is not only a special feature of the magnetization curves in this compound, sharp additional singularities are seen at several critical fields for $T < T_N$. As shown in Fig. 4(left panel), the derivatives $dM/dB$ of the magnetization curves taken at $T < T_N$ highlight these features, i.e. peaks at $B_1$ and $B_2$, a sharp change of slope at $B_3$ and full saturation at $B_4$. The magnetization curves taken at low temperatures can be modeled quantitatively if one presumes the magnetic moments of $Cu^I$ and $Cu^{II}$ ions forming separate magnetic subsystems and experiencing the spin-flop and spin-flip transitions independently. The satisfactory fit of the experimental data is obtained ascribing $B_1$ and $B_4$ to spin-flop and spin-flip transitions in the $Cu^{II}$ subsystem and $B_2$ and $B_3$ to spin-flop and spin-flip transitions in the $Cu^I$ subsystem, correspondingly. In the mean-field approximation $B_{flop} = (2B_A B_E - B_A^2)^{1/2}$ and $B_{flip} = B_E$, where $B_A$ is the anisotropy field and $B_E$ is the exchange field. Using our experimental values for the critical fields we arrive at $B_A^I = 0.8T$, $B_E^I = 3.15T$ and $B_A^{II} \sim 0.3T$, $B_E^{II} \sim 5.2T$.

Overall, this points to a possible decoupling in the magnetic subsystem of $Ba_3Cu_3In_4O_{12}$. At increasing temperatures, the M(B) curves smooth out reaching the standard Brillouin function, but the progressively smeared anomalies at the spin-flop-like and spin-flip-like transitions persist. This is summarized in the magnetic phase diagram which is constructed from the magnetization M and specific heat Cp data, as shown in Fig. 4(right panel). The $B_1$ and $B_2$ lines in this diagram correspond to spin-flop transitions in $Cu^{II}$ and $Cu^I$ subsystems (note, the ~ 2:1 ratio in magnitude of subsequent singularities, as shown in the inset to Fig. 4(left panel)), while the $B_3$ and $B_4$ lines denote spin-flip transitions in $Cu^I$ and $Cu^{II}$ subsystems.

The decoupling in the magnetic subsystem in $Ba_3Cu_3In_4O_{12}$ is only possible for an orthogonal spin arrangement of the copper ions. Such highly unusual magnetic structure can be formed to avoid heavy frustration of exchange interactions in $Ba_3Cu_3In_4O_{12}$. The frustration of ferromagnetic nearest-neighbor $Cu^I$–O–$Cu^{II}$ interactions by antiferromagnetic next-nearest-neighbor $Cu^{II}$–O–O–$Cu^{II}$ interactions is not the only factor influencing the formation of three-dimensional long-range order in $Ba_3Cu_3In_4O_{12}$. The intrachain $Cu^{II}$–$Cu^{II}$ interaction within the columns is intrinsically frustrated itself due to the tetrahedral arrangement of $Cu^{II}$ ions. Moreover, the interchain interaction within the $Cu^{II}$ subsystem is frustrated, too. As shown in Fig. 5, the presumably decoupled copper subsystem in $Ba_3Cu_3In_4O_{12}$ consists of a three-dimensional rectangular $Cu^I$ network (left panel) embedded between mutually orthogonal $Cu^{II}$ two-dimensional Shastry-Sutherland layers (right panel), resulting in a three-dimensional heavily frustrated Shastry-Sutherland system[26]. In generalized Shastry-Sutherland model the dimers on neighboring layers have no direct coupling, but couple to an intermediate spin between them[27].

For the Shastry-Sutherland model, it is clear that the system will exhibit long-range Neel order at predominance of interdimer interaction, and will be in the short-range dimer state at prevalence of intradimer interaction. Recent studies suggest that there exists an intermediate phase between the Neel phase and the dimer phase[28]. While the nature of a possible intermediate state in three-dimensional Shastry-Suthertland model remains an open question, one possible candidate could be an orthogonal antiferromagnetic network[29].

## CONCLUSION

In our view, the ground state in $Ba_3Cu_3In_4O_{12}$ is comprised of three virtually independent orthogonal subsystems, as shown in Fig. 5(bottom panel). No frustration is present in this arrangement. The weakness of relevant exchange interactions results in rather low value of $B_S$, while the absence of frustration in the orthogonal spin arrangement leads to a relatively high $T_N$. In this structure, favored by magnetocrystalline anisotropy, pseudodipolar and Dzyaloshinskii-Moriya interactions, quantum fluctuations provide the coupling between mutually orthogonal magnetic subsystems resulting in an impressive "order-by-disorder" effect. The Schottky-type anomaly in the specific heat may define the energy of stabilization of the quantum ground state in this system. The low magnetic field necessary to saturate the magnetization in $Ba_3Cu_3In_4O_{12}$ tentatively indicates the proximity of this system to a quantum critical point separating orthogonal and collinear phases.

While the neutron scattering experiment on $Ba_3Cu_3In_4O_{12}$ is hampered by high absorption cross – section of indium, the recently performed neutron diffraction study on isostructural compound $Ba_3Cu_3Sc_4O_{12}$ has revealed complex magnetic order at low temperatures which is greatly influenced by external magnetic field[30]. The analysis of available data on $Ba_3Cu_3Sc_4O_{12}$ allows excluding ferromagnetic "paper-chain" model for this compound. There are no magnetic reflections corresponding to this arrangement, while the magnetic reflection with propagation vector **k** = [010] is compatible with orthogonal antiferromagnetic network. It does not allow, however, distinguish between collinear and orthogonal antiferromagnetic structures. Our measurements of thermodynamic properties of $Ba_3Cu_3Sc_4O_{12}$ ($T_N$ = 16K) show its full similarity to $Ba_3Cu_3In_4O_{12}$ making, therefore, these compounds new family of non-collinear antiferromagnets.

Concluding, the magnetic subsystem in $Ba_3Cu_3In_4O_{12}$ seems to be frustrated in every sector regarding intrachain and interchain interactions. The formation of an orthogonal spin structure lifts all these multiple frustrations. In accordance with the orthogonal arrangement of copper magnetic moments the $Cu^I$ and $Cu^{II}$ subsystems are to be decoupled and the $Cu^{II}$ subsystem decouples into two equivalent sub-subsystems itself. This is in agreement with the observation of two distinct spin-flop-like and two distinct spin-flip-like transitions of different magnitudes. In case of exclusion of the isotropic superexchange interaction in our consideration, other weak, basically anisotropic interactions masked in helix and collinear antiferromagnets manifest themselves, making $Ba_3Cu_3In_4O_{12}$ a new and excellent playground to study the physics of low-dimensional and frustrated quantum spin systems.

**ACKNOWLEDGEMENTS**

We acknowledge the support of the present work by Deutsche Forschungsgemeinschaft Grants DFG 486 RUS 113/982/0-1 and WO 1532/3-1, Russian Foundation for Basic Research Grants RFBR 09-02-91336, 10-02-00021, 11-02-00083.

1. M. Jaime, V.F. Correa, N. Harrison, et. al., Phys. Rev. Lett., **93**, 087203 (2004).
2. M. Hase, I. Terasaki, K. Uchinokura, Phys. Rev. Lett. **70**, 3651 (1993).
3. J. Darriet, L.P. Regnault, Solid State Comm., **86**, 409 (1993).
4. H. Kageyama, K. Yoshimura, R. Stern, et. al., Phys. Rev. Lett., **82**, 3168 (1999).
5. T. Masuda, A. Zheludev, B. Roessli, et al., Phys. Rev. B **72**, 014405 (2005).
6. M. Enderle, C. Muckerjee, B. Fäk, et. al., Europhys. Lett., **70**, 237 (2005).
7. Y. Tarui, Y. Kobayashi, M. Sato, J. Phys. Soc. Jpn., **77**, 043703 (2008).
8. S.-L. Drechsler, J. Richter, R. Kuzian, et al., J. Magn. Magn. Mater. **316**, 306 (2007).
9. R. Bursill, G.A. Gehring, D.J.J. Farnell, et al., J. Phys.: Condens. Matter **7**, 8605 (1995).
10. R. Sachidanandam, T. Yildirim, A.B. Harris, et. al., Phys. Rev. B **56**, 260 (1997).
11. L.X. Hayden, T.A. Kaplan, S.D. Mahanti, Phys. Rev. Lett., **105**, 047203 (2010).
12. D. Petitgrand, S.V. Maleev, Ph. Bourges, A.S. Ivanov, Phys. Rev. B **59**, 1079 (1999).
13. S. Tornow, O. Entin-Wohlman, A. Aharony, Phys. Rev. B, **60**, 10206 (1999).
14. L.E. Aleandri, H.G. von Schnering, Journal of the Less-Common Metals, **156**, 181 (1989).
15. D.H. Gregory, P.R. Mawdsley, S.J. Barker, et. al., Journal of Materials Chemistry **11**, 806 (2001).
16. A.L. Kharlanov, N.R. Khasanova, M.V. Paramonova, et. al., Zh. Neorg. Khim. (in Russian), **35**, 3067 (1990).

17. G.A. Bain, J.F. Berry, J. Chem. Educ. **85**, 532 (2008).
18. R. Klingeler, B. Büchner, S.-W. Cheong, H. Hücker, Phys. Rev. B 72, 104424 (2005).
19. M. E. Fisher, Phil. Mag., **7**, 1731 (1962).
20. S.-L. Drechsler, J. Richter, R. Kuzian, et al. J. Magn. Magn. Mater **316**, 306 (2007).
21. M. Schmitt, J. Malek, S.-L. Drechsler, H. Rosner, Phys. Rev. B **80**, 205111 (2009).
22. W.E.A. Lorenz, R.O. Kuzian, S.-L. Drechsler, et al. Europhys. Lett. **88**, 37002 (2009).
23. H.J. Schulz, Phys. Rev. Lett., **77**, 2790 (1996).
24. V.Yu. Irkhin, A.A. Katanin, Phys. Rev. B, **55**, 12318 (1997).
25. N. Buettgen, H.-A. Krug von Nidda, L.E. Svistov et. al., Phys. Rev. B, **76**, 014440 (2007).
26. B.S. Shastry, B. Sutherland, Physica B, **108**, 1069 (1981).
27. S. Chen, H.Buettner, Eur. Phys. J. B, **29**, 15 (2002).
28. W. Zheng, J. Oitmaa, C.J. Harner, Phys. Rev. B, **65**, 014408 (2001).
29. D. Carpentier, L. Balents, Phys. Rev. **B, 65**, 024427 (2001).
30. A.V. Mahajan, B. Koteswara Rao, J. Bobroff, Bull. Amer. Phys. Soc., **54**, 06.14.7 (2009); V. Siruguri, P.D. Babu, A.V. Pimpale, et al., UGC-DAE CSR Bulletin **19**, 5 (2010).

Table 1. Basic parameters of the magnetic subsystem in selected quasi-one-dimensional compounds.

| Compound | $T_N$, K | J, K | $J_1$, K | $J_2$, K | $B_S$, T | Comment |
|---|---|---|---|---|---|---|
| LiCuVO$_4$ | 2.3 | - 18 | 49 | 9 | 46.2 | helix chain (Ref. 25) |
| Li$_2$CuO$_2$ | 9.4 | - 228 | 76 | 9 | 55.4 | ferromagnetic chain (Ref. 22) |
| Ba$_3$Cu$_3$In$_4$O$_{12}$ | 12.7 | - 84 | 8 | 2 | 5.2 | paper chain (present work) |

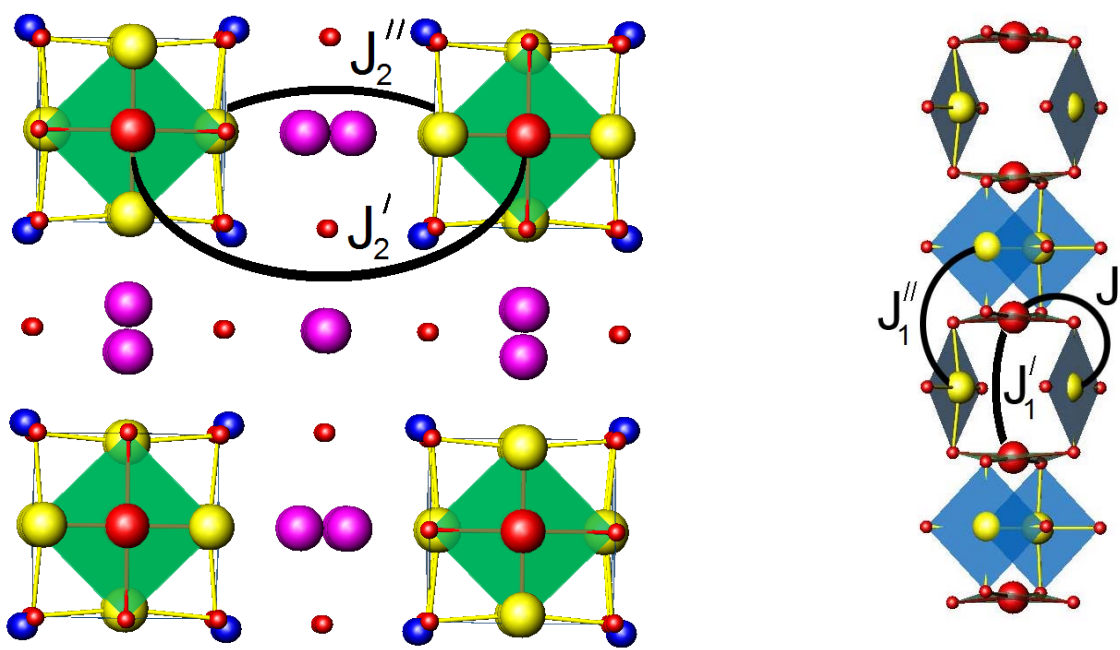

Fig. 1. The Ba$_3$Cu$_3$In$_4$O$_{12}$ structure projected along the c- axis. Large and medium isolated spheres represent the Ba$^{2+}$ and the In$^{3+}$ ions. Small spheres are the O$^{2-}$ ions. The CuO$_4$ units are shown in polyhedral representation. The $J_2'$ and $J_2''$ arcs mark Cu$^I$ – Cu$^I$ and Cu$^{II}$ – Cu$^{II}$ interchain exchange interactions, respectively (left panel). The paper – chain column consists of vertex – sharing buckled Cu$^I$O$_4$ (horizontal) and concave Cu$^{II}$O$_4$ (vertical) units. The J arc marks nearest-neighbor Cu$^I$ – Cu$^{II}$ exchange interaction. The $J_1'$ and $J_1''$ arcs mark Cu$^I$ – Cu$^I$ and Cu$^{II}$ – Cu$^{II}$ next-nearest-neighbor intrachain exchange interactions, respectively (right panel).

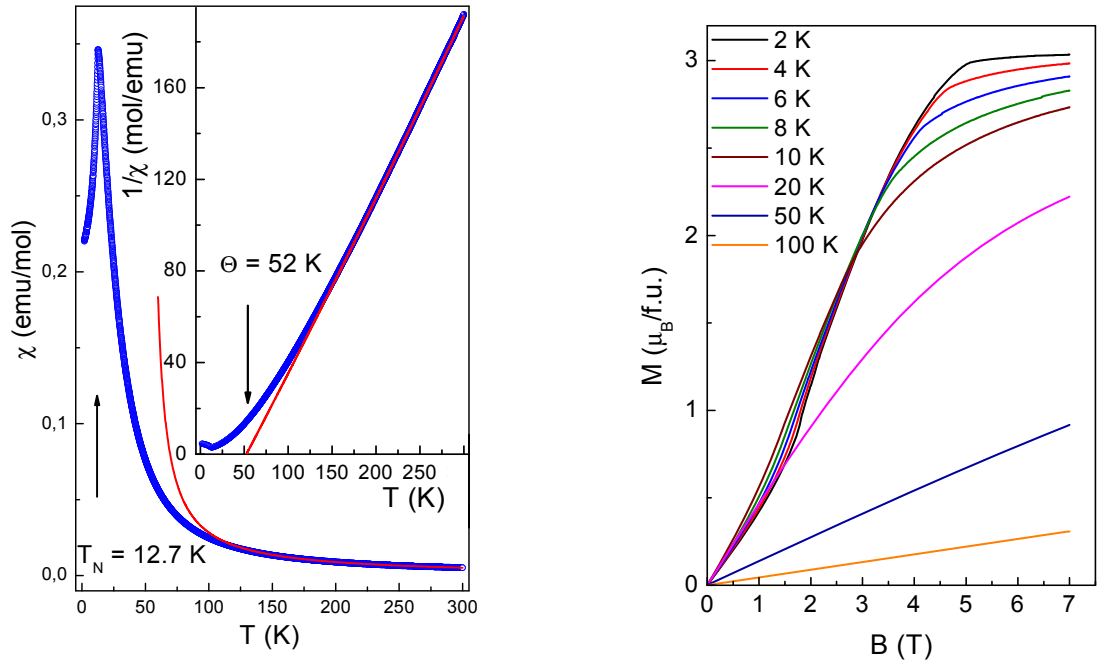

Fig. 2. Magnetic susceptibility $\chi(T)$ of $Ba_3Cu_3In_4O_{12}$ taken at B = 0.1 T. The inset shows the temperature dependence of $\chi^{-1}$. The solid lines represent the Curie – Weiss fit of the high – temperature data (left panel). The field dependences of magnetization in $Ba_3Cu_3In_4O_{12}$. The fitting of M vs. B curves by modified Brillouin function $B_{1/2}$ as described in the text gives number N of ferromagnetically correlated spins in paramagnetic state as 97 at 100 K, 151 at 50 K, and 208 at 20 K (right panel).

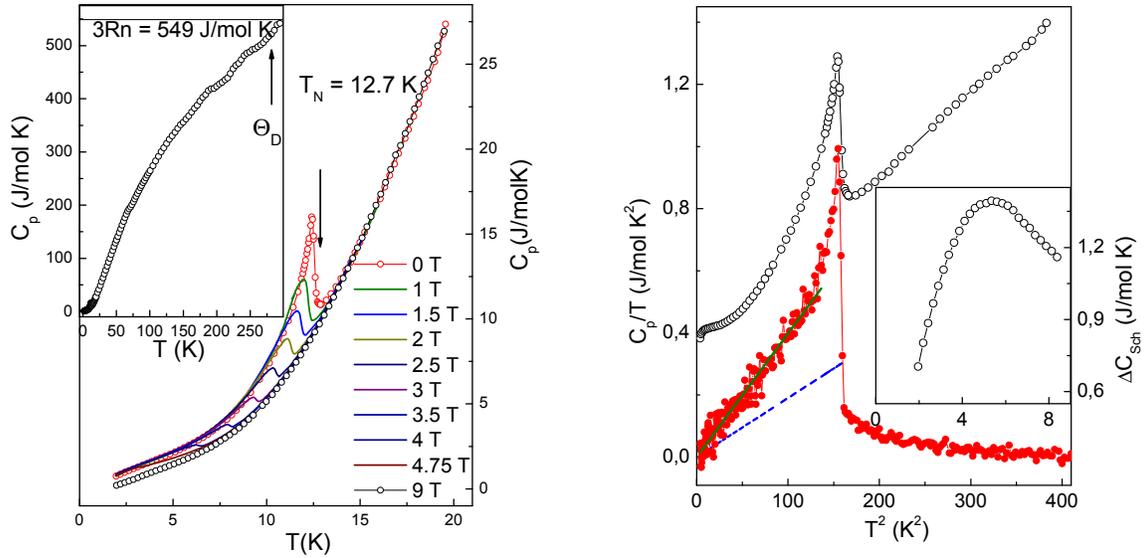

Fig. 3. Temperature dependences of the specific heat Cp in $Ba_3Cu_3In_4O_{12}$ taken under various magnetic fields. Inset: Cp(T) measured up to room temperature. The deviations from a "smooth" behavior at elevated temperatures are due to the well-known influence of the Apiezon-N grease used to fix the sample in the specific heat puck of the PPMS device (left panel). The specific heat data in $Ba_3Cu_3In_4O_{12}$ in Cp/T vs. $T^2$ representation (open symbols). The normalized Fisher's specific heat (solid symbols) is calculated from the experimental χ(T) data. The dotted and solid lines represent the phonon and magnon contributions at low temperatures, correspondingly. Inset: extra contribution to the specific heat at low temperatures of Schottky-type (right panel).

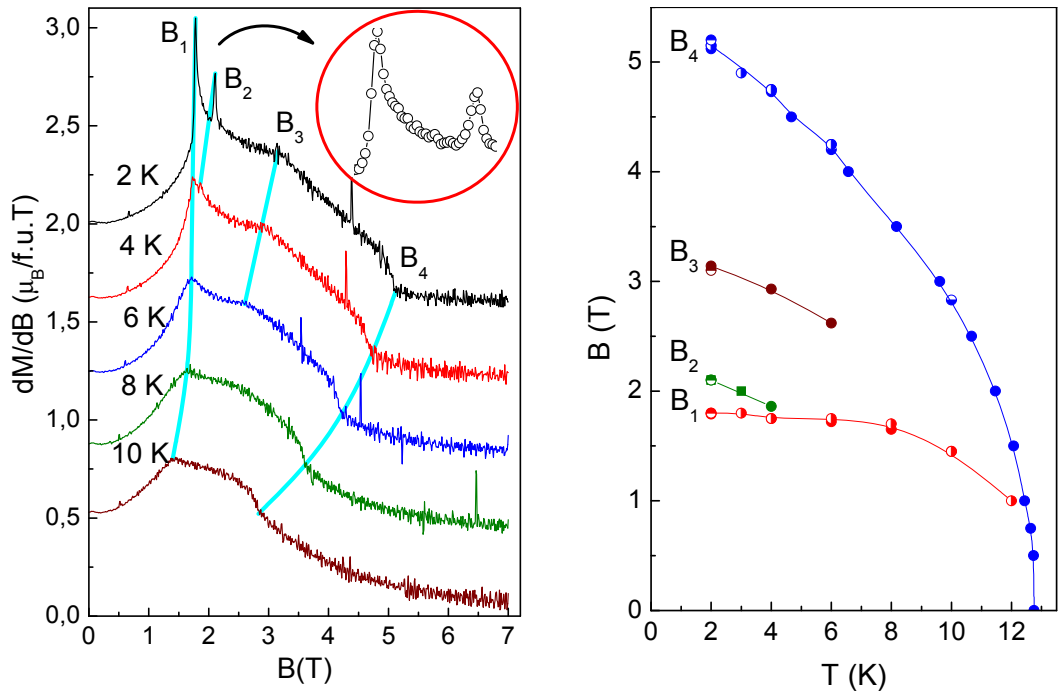

Fig. 4. The derivatives of the magnetization dM/dB of $Ba_3Cu_3In_4O_{12}$. The subsequent dM/dB curves are shifted with respect to each other. Solid lines are guides to the eye to follow the subsequent spin – flop and spin – flip transitions. The inset represents the enlarged region of spin – flop transitions (left panel). The B – T magnetic phase diagram in $Ba_3Cu_3In_4O_{12}$ as compiled from the specific heat Cp and magnetization M data (right panel).

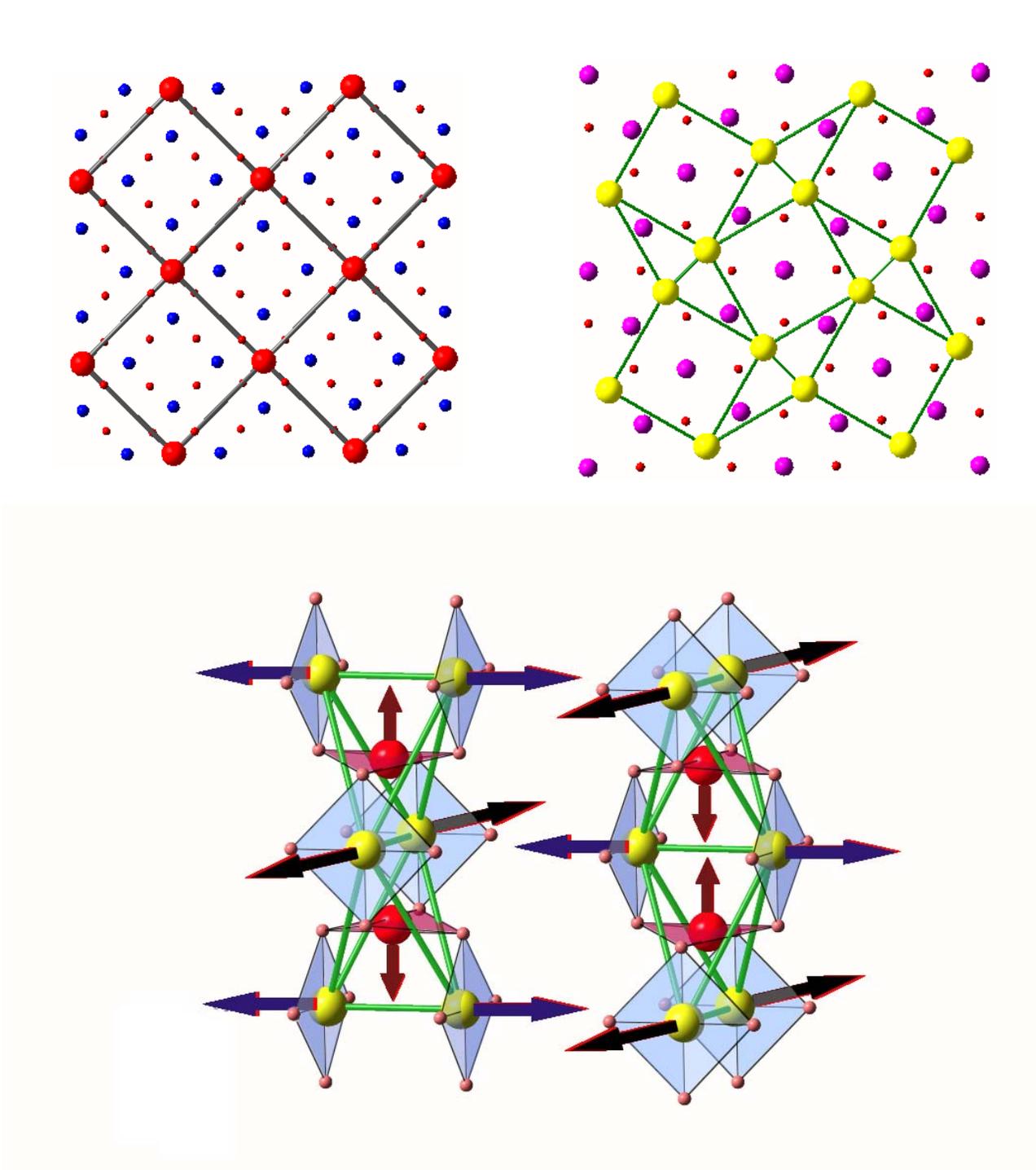

Fig. 5. The structure of Cu$^I$ layers (left panel) and Cu$^{II}$ layers (right panel) in the *ab* plane in Ba$_3$Cu$_3$In$_4$O$_{12}$. The Cu$^{2+}$ ions are interconnected by solid lines. The large, medium and small size isolated spheres represent Ba$^{2+}$, In$^{3+}$ and O$^{2-}$ ions, respectively. Note, that the adjacent Cu$^{II}$ layers are rotated by 90° with respect to each other along the *c* - axis. The orthogonal spin arrangement of copper magnetic moments as possible ground state of generalized three – dimensional Shastry – Sutherland network in Ba$_3$Cu$_3$In$_4$O$_{12}$ (bottom panel).